\documentclass[nofootinbib,a4paper,aps,prd,10pt,superscriptaddress,showkeys,twocolumn]{revtex4}

\usepackage{graphicx}
\usepackage{amsfonts}
\usepackage{amssymb}
\usepackage{amsmath}
\usepackage{natbib}
\usepackage{float}
\usepackage{dcolumn}
\usepackage{bm}
\usepackage{latexsym,color}

\def\nat{Nature}

\def\prd{Phys. Rev. D}

\def\mnras{Mon. Not. Roy. Astr. Soc.}
\def\apj{Astrophys. J.}

\def\pasj{Publications of the Astronomical Society of Japan }

\newcommand{\bea}{\begin{eqnarray}}
\newcommand{\eea}{\end{eqnarray}}
\newcommand{\be}{\begin{equation}}
\newcommand{\ee}{\end{equation}}

\begin{document}

\title{Accretion disk luminosity for black holes surrounded by dark matter \\ with anisotropic pressure}

\author{Kuantay~\surname{Boshkayev}}
\email[]{kuantay@mail.ru}
\affiliation{National Nanotechnology Laboratory of Open Type,  Almaty 050040, Kazakhstan.}
\affiliation{%
Al-Farabi Kazakh National University, Al-Farabi av. 71, 050040 Almaty, Kazakhstan.
}
\affiliation{%
Department of Physics, Nazarbayev University, Kabanbay Batyr 53, 010000 Nur-Sultan, Kazakhstan.
}

\author{Roberto Giamb\`o}
\email[]{roberto.giambo@unicam.it}

\affiliation{Scuola di Scienze e Tecnologie, Universit\`a di Camerino, Via Madonna delle Carceri 9, 62032 Camerino, Italy.}

\affiliation{Istituto Nazionale di Fisica Nucleare, Sezione di Perugia, Via Alessandro Pascoli 23c, 06123 Perugia, Italy.}

\author{Talgar~\surname{Konysbayev}}
\email[] {talgar_777@mail.ru}
\affiliation{National Nanotechnology Laboratory of Open Type,  Almaty 050040, Kazakhstan.}

\affiliation{%
Al-Farabi Kazakh National University, Al-Farabi av. 71, 050040 Almaty, Kazakhstan.
}

\author{Ergali~\surname{Kurmanov}}
\email[]{kurmanov.yergali@kaznu.kz}
\affiliation{National Nanotechnology Laboratory of Open Type,  Almaty 050040, Kazakhstan.}

\affiliation{%
Al-Farabi Kazakh National University, Al-Farabi av. 71, 050040 Almaty, Kazakhstan.
}

\author{Orlando~\surname{Luongo}}
\email[]{orlando.luongo@unicam.it}

\affiliation{%
Al-Farabi Kazakh National University, Al-Farabi av. 71, 050040 Almaty, Kazakhstan.
}

\affiliation{Scuola di Scienze e Tecnologie, Universit\`a di Camerino, Via Madonna delle Carceri 9, 62032 Camerino, Italy.}

\affiliation{%
Dipartimento di Matematica, Universit\`a di Pisa, Largo B. Pontecorvo 5, Pisa, 56127, Italy.
}

\author{Daniele~\surname{Malafarina}}
\email[]{daniele.malafarina@nu.edu.kz}
\affiliation{%
Department of Physics, Nazarbayev University, Kabanbay Batyr 53, 010000 Nur-Sultan, Kazakhstan.
}

\author{Hernando~\surname{Quevedo}}
\email[]{quevedo@nucleares.unam.mx}
\affiliation{%
Al-Farabi Kazakh National University, Al-Farabi av. 71, 050040 Almaty, Kazakhstan.
}
\affiliation{%
Instituto de Ciencias Nucleares, Universidad Nacional Aut\'onoma de M\'exico, Mexico.
}
\affiliation{%
Dipartimento di Fisica and ICRA, Universit\`a di Roma “La Sapienza”, Roma, Italy.
}

\date{\today}

\begin{abstract}
We investigate the luminosity of the accretion disk for a static black hole surrounded by dark matter with anisotropic pressure. We calculate all basic orbital parameters of test particles in the accretion disk, such as angular velocity, angular momentum, energy and radius of the innermost circular stable orbit as functions of the dark matter density, radial pressure and anisotropic parameter, which establishes the relationship between the radial and tangential pressures. We show that the presence of dark matter with anisotropic pressure makes a noticeable difference in the geometry around a Schwarzschild black hole, affecting the radiative flux, differential luminosity and spectral luminosity of the accretion disk.
\end{abstract}



\maketitle

\section{Introduction}

The common disk-like particle flow lying around compact objects, dubbed accretion disk, permits direct observations of the material orbiting in the gravitational field of a given astronomical object \cite{agn1, agn2,abramowicz}. In particular, the corresponding spectra are commonly observed, providing  information about the characteristics of the  central object that has led to the formation of the accretion disk itself \cite{1988ApJ...332..646A}.  For extreme compact objects the relativistic effects are clearly non-negligible and so the accretion disk luminosity of objects must be framed through the Einstein field equations. The central compact object, namely the accretor, can be modeled through particular spacetimes, with given symmetries \cite{shapirobook, 2002NuPhB.626..377T}. In this picture, one can investigate black holes, white dwarfs, neutron stars, quasars, radio galaxies, X-ray binaries
but also more exotic, hypothetical objects, e.g. boson stars \cite{2002NuPhB.626..377T,2006PhRvD..73b1501G} or gravastars \cite{2004PNAS..101.9545M, 2006CQGra..23.2303D}.

The accretion region is characterized by the growth in mass due to the gravitational attraction. Thus, to fully-describe those disks, one requires an exterior configuration and then solves the equations for hydrodynamic equilibrium for the interior matter content \cite{ht1968,2017LRR....20....7P}. Frequently, ``exotic'' matter contributions seem to be needed to describe hypothetical objects that are massive and compact, in order to fulfill stability criteria. Similar examples can often be found for wormholes and cosmological backgrounds \cite{2021EPJP..136..167C,2006IJMPD..15.1753C,LM}, and likely   exotic matter could indicate a possible signature of general relativity's breakdown  \cite{2019IJMPD..2830016C,2019PhRvD.100h3513D}.

In this work, we consider a static spherically symmetric configuration composed of a central black hole surrounded by a dark matter envelope. The gravitational field in the vacuum region around the black hole is described by the exterior Schwarzschild space-time while the corresponding dark matter distribution, located at a given distance from the black hole,
and its properties, is described by making use of the Tolman-Oppenheimer-Volkoff (TOV) equations.

We assume that the dark matter envelope does not interact with the baryonic matter of the accretion disk that is located within the envelope itself.
We then apply the theory of black hole accretion developed in Ref.~\cite{novikov1973} for astrophysical black hole candidates, in order to model the emitted spectrum from the accretion disk. In particular, we aim to test the consequences of two main assumptions:
(i) dark matter in endowed with a non-vanishing radial pressure term entering the TOV equations, namely $P_r(r)=P(r)$ and (ii) second,  we assume the energy momentum tensor to be anisotropic, leading to an additional pressure term, which is interpreted as a non vanishing tangential pressure, $P_\theta(r)$. We physically motivate these two choices and characterize the dark matter distribution accordingly, by computing the difference $P_\theta(r)-P(r)$.

It should be stressed that the theory of anisotropic fluids is well known in the literature. In particular, it was shown that anisotropic fluids may be geodesic in general relativity in Ref.~\cite{2002JMP....43.4889H}. A general study of spherically symmetric dissipative anisotropic fluids is given in Ref.~\cite{2004PhRvD..69h4026H}. Exact static spherically symmetric anisotropic solutions of the field equations are obtained and analyzed in Refs.~\cite{1982PhRvD..26.1262B,2008PhRvD..77b7502H}. Anisotropic stars in general relativity and their mass-radius relations are computed in Ref.~\cite{2003RSPSA.459..393M}. In this work we instead study the effects of dark matter with anisotropic pressures on
test particles in the accretion disk present within the dark matter and the spectra of the disk.
In particular we compare the motion of particles and accretion disk's spectra with the cases of isotropic dark matter and a Schwarzschild black hole in vacuum.

The paper is organized as follows: in Sect.~\ref{sez2}, we describe a configuration
that consists of a black hole surrounded by a dark matter distribution. We then introduce the static line element with anisotropic energy  momentum tensor containing tangential pressure $P_{\theta}$ used to describe the dark matter envelope.
Afterwards, we review the definitions of flux, differential luminosity, and spectral luminosity as presented in the  Novikov-Page-Thorne (NPT) model. In Sect.~\ref{sez3}, we solve the Tolman-Oppenheimer-Volkoff (TOV) equations and calculate the metric functions fulfilling boundary conditions in the different regions of the space-time.
Then we compute the angular velocity, energy, and angular momentum of the configuration and plot the
modeled flux and luminosity spectrum for various values of the parameter related to the dark matter anisotropy. Implications of the model for astrophysical black hole candidates are then discussed in in Sect.~\ref{sez4}.



\section{Black hole surrounded by anisotropic dark matter}\label{sez2}

In the following we investigate a system composed of a static black hole with a dark matter envelope around it, where the dark matter is introduced \emph{only} with the purpose of modifying the geometry around the black hole. Since the black hole represents the accretor, one can split the mass profile of the overall configuration into three regions,
\begin{equation}\label{eq:massprof2}
    M(r)=\left\{
                \begin{array}{lll}
                  M_{BH},  \quad \qquad \qquad \qquad r_g < r \leq r_b ,\\
                  M_{BH}+M_{DM}(r), \quad \quad r_b \leq r \leq r_s ,\\
                 M_{BH}+M_{DM}(r_s), \, \, \, \quad r_s \leq r ,
                \end{array}
              \right.
\end{equation}
where $r_g=2 M_{BH}$ is the gravitational radius and $M_{BH}$ is the mass black hole, whereas $r_b$ and $r_s$ are the inner and outer edges (radii) of the dark matter envelope. In particular, $r_b$ corresponds to the boundary that separates the inner vacuum region
from the outer distribution of dark matter.
Accordingly, the above configuration can be described as follows:
\begin{itemize}
\item the core is modeled by the accretor in the form of  a black hole. Its mass, $M_{BH}$, is a free parameter of our model;
\item the black hole is surrounded by a dark matter shell that extends from a radius $r_b$ up to the radius $r_s$;
\item at $r_s$, the dark matter mass reaches its maximum value
$M_{DM}(r_s)$ and beyond $r_s$ we assume vacuum.
\end{itemize}

%
%
To model the dark matter distribution in the  shell $r\in[r_b,r_s]$, we assume  an exponential sphere profile of the form
 \color{black}
\begin{equation}\label{eq:den}
\rho(r) = \rho_0  e^{-\frac{r}{r_0}}, \quad r\geq r_b,
\end{equation}
where $\rho_0$ is the dark matter density at $r=0$ and  $r_0$ is the scale radius. The exponential density profile was introduced in Ref.~\cite{2013PASJ...65..118S} to explain the rotation curve in the bulge of the Milky Way Galaxy. Indeed it showed a better fit of the observational data with respect to the widely adopted de Vaucouleurs law in the inner part of the galaxy.

As a consequence, assuming spherical symmetry, the dark matter mass profile is given by
\begin{eqnarray}\label{eq:mdmprof}
M_{DM}(r)&=&\int_{r_b}^r 4 \pi  \tilde{r}^2 \rho (\tilde{r}) \, d\tilde{r},
\end{eqnarray}
which yields
\begin{eqnarray}
    M_{DM}(x)&=&8 \pi r_{0}^3 \rho_{0}\Bigg[e^{-x_b} \left(1+x_b+\frac{x_b^2}{2}\right) \nonumber\\ &&-e^{-x}\left(1+x+\frac{x^2}{2}\right)\Bigg],
\end{eqnarray}
for $r>r_b$, where we have substituted $x=r/r_0$ and $x_b=r_b/r_0$. For vanishing $r_b$, the profile \eqref{eq:mdmprof} reduces to the one obtained in \cite{2013PASJ...65..118S}.

\subsection{TOV equations with anisotropic pressure}

To describe the physical properties of the system in Eq. \eqref{eq:massprof2}, we consider the spherical symmetric line element
\begin{equation}\label{eq:le}
d s^2=e^{N(r)} d t^2 - e^{\Lambda(r)} d r^2 - r^2 \left(d \theta^2 + \sin^2 \theta d\varphi^2\right),
\end{equation}
where, as usual, we take $(t,r,\theta,\varphi)$ as time and spherical coordinates, respectively, while $N(r) $ and $\Lambda(r)$ represent the unknown metric functions.

The energy-momentum tensor is given by
\begin{equation}\label{T}
T^{\alpha\beta} = (\rho + P_\theta) u^\alpha u^\beta -P_\theta  g^{\alpha\beta} +(P - P_\theta)\chi^\alpha \chi^\beta  \ ,
\end{equation}
where $u^\alpha u_\alpha=1=-\chi^\alpha\chi_\alpha$, $u^\alpha \chi_\alpha=0$, $u^\alpha=e^{-N/2}\delta^\alpha_0$ is the four-velocity and $\chi^\alpha=e^{-\Lambda/2}\delta^\alpha_1$ is a unit space-like vector in the radial direction, with $\delta^{\alpha}_i$ the Kronecker symbol \cite{2021JPhCS1816a2025R}.
Using Einstein's equations for the line element \eqref{eq:le} and the energy-momentum tensor\eqref{T} we obtain the following expressions \cite{1996PhRvD..54.4862F}
\begin{eqnarray}
\frac{d P(r)}{d r}&=&-(\rho(r) +P(r)) \frac{M(r)+4 \pi r^3 P(r)}{r(r-2 M(r))}\label{pressurediff} \\ \nonumber
&& +\frac{2}{r}\left(P_{\theta}(r)-P(r)\right),\\
\frac{d N(r)}{dr}&=&
2\,\frac{M(r)+4\pi r^3 P(r)}{r(r-2M(r))},
\end{eqnarray}
which generalize the TOV equations to the case of anisotropic pressures and relate the density, $\rho(r)$, with the radial pressure, $P(r)$, and the tangential term, $P_{\theta}(r)$.
In general this is a system of two equations in four unknown, and therefore two functions must be provided in order to close it. Typically this is done by specifying the equations of state that relate the pressures $P_{\theta}(r)$ and $P(r)$ to the density.
In our model, however, $\rho(r)$ is the dark matter density given by Eq.~\eqref{eq:den}, whereas $M(r)$ is given by Eq.~\eqref{eq:massprof2}; and therefore, these quantities are no longer unknown functions to be determined. Notice,
however, that one unknown function remains to be specified. This function can be taken to be the pressure difference $P_{\theta}(r)-P(r)$, appearing in Eq. \eqref{pressurediff}, which is not known \emph{a priori}.

A possible viable strategy to employ has been discussed in Ref.~\cite{2011CQGra..28b5009H,2015PhRvD..91d4040F}, where the \emph{anisotropy function},  $\Delta(r)$, has been introduced as
\begin{equation}
 \label{eq:delta}
\Delta(r)\equiv P_{\theta}(r)-P(r)=\alpha \mu(r) P(r)\,,
\end{equation}
where $\alpha$ is a free constant, physically interpreted as the anisotropy parameter, while $\mu$ is the \emph{compactness} of the system  defined by
\begin{equation}
 \mu(r)=\frac{2 M(r)}{r}=1-e^{-\Lambda}\,.
\end{equation}
In general,  $\mu$ indicates the strength of the gravitational field. If $\mu\ll1$ the field is weak, if $\mu\sim1$ the field is strong. An interesting characteristics of Eq.~\eqref{eq:delta} is that $\mu$ guarantees the required vanishing of the anisotropy of pressures at $r=0$ (notice that typically $M(r)\sim r^3$ close to the center). Moreover in the weak field limit the anisotropy of pressure is not expected to be important. In addition, this anzatz makes sure that the tangential pressure vanishes at the surface of the object \cite{2011CQGra..28b5009H}, in our case at the surface of dark matter envelope.

Notice that according to Eq.~\eqref{eq:delta}, $\alpha<0$ corresponds to $P_\theta<P$, $\alpha=0$ corresponds to the isotropic case $P_\theta=P$ and $\alpha>0$ corresponds to $P_\theta>P$.
The case of a static black hole surrounded by a dark matter envelope with isotropic pressures, i.e. $\alpha=0$, has been studied in detail by some of us in Ref.~\cite{2020MNRAS.496.1115B} within a model that closely follows the one considered here.

Furthermore, to analyze our model, we should now  establish the boundary conditions between inner and outer solutions.

\subsection{Boundary conditions}

As stated above, the innermost region is given by a black hole vacuum solution with the event horizon located at $r=r_g$, while the exterior region corresponds to the dark matter distribution {extending from a radius $r_b>r_g$ to an outer radius $r_s>r_b$. The values of density, pressures and metric functions at the boundary are determined from} $\rho(r_b), P(r_b)$ and $N(r_b)$. They can be computed immediately, obtaining

\begin{eqnarray}
\rho(r_b)&=&\rho_b=\rho_0 \, e^{-\frac{r_b}{r_0}}, \\
P(r_b)&=&P_b, \\
N(r_b)&=&N_b=\ln{\left(1-\frac{r_g}{r_b}\right)}\,.
\end{eqnarray}
Consequently,  the unknown metric functions, $N(r)$ and $\Lambda(r)$, are evaluated as
\begin{equation}\label{eq:bc}
    e^{N (r)}=\left\{
                \begin{array}{lll}
                  1-\dfrac{r_g}{r}, \quad \qquad r_g < r \leq r_b ,\\
                  e^{N_{r} (r)}, \, \qquad \quad r_b \leq r \leq r_s ,\\
                   1-\dfrac{2 M(r_s)}{r}, \, \, \quad r_s \leq r ,
                \end{array}
              \right.
\end{equation}
 and
\begin{equation}\label{eq:bc22}
    e^{\Lambda (r)}=\left\{
                \begin{array}{lll}
                  \left(1-\dfrac{r_g}{r}\right)^{-1},\qquad \quad r_g < r \leq r_b ,\\
                  \left(1-\dfrac{2 M(r)}{r}\right)^{-1}, \, \quad r_b \leq r \leq r_{s} ,\\
                  \left(1-\dfrac{2 M(r_s)}{r}\right)^{-1}, \, \, \, \, \, r_s \leq r .
                \end{array}
              \right.
\end{equation}
where $N_{r} (r)$ is simply function $N(r)$ in the interval $r\in[r_b,r_s]$ which must be numerically evaluated from the TOV equations fulfilling the corresponding boundary conditions.

It is worth noticing that if we follow this consolidate procedure for matching different spacetimes by imposing continuity of the first and second fundamental forms, the dark matter pressure should vanish when $r=r_b$. Therefore the condition $P(r_b)=P_b$ leads to a non-continuous matching, since  the first derivatives of the metric show a jump at the boundary. However, there is a natural physical explanation for the discontinuity. The common interpretation is to assume the presence of a massive surface layer at $r_b$ for which the three dimensional energy momentum can be evaluated from the matching conditions, as discussed in  Refs.~\citep{1966NCimB..44....1I,1967NCimB..48..463I}.


\subsection{Radiative flux and spectral luminosity}
Bearing in mind the above results, we can now investigate the flux and spectral luminosity produced by an accretion disk in the geometry with the above proposed ansatz. The disk extends from an inner edge $r_i$ which is usually taken as the innermost stable circular orbit (ISCO) for test particles, $r_i=r_{ISCO}$. To this end, we follow the simple approach  proposed by Novikov-Thorne and Page-Thorne in Refs.~\cite{novikov1973, page1974} and write the radiative flux $\mathcal{F}$ as

\begin{equation}
 \label{eq:flux}
\mathcal{F}(r)=-\frac{\dot{{\rm m}}}{4\pi \sqrt{g}} \frac{\Omega_{,r}}{\left(E-\Omega L\right)^2 }\int^r_{r_{i}} \left(E-\Omega L\right) L_{,\tilde{r}}d\tilde{r}.
\end{equation}
The above quantity depends upon  $\dot{{\rm m}}$, { i.e.}, the disk mass accretion rate, which is unknown. In the simplest case, we can take it as constant and we can set $\dot{{\rm m}}=1$, which is equivalent to considering the normalized flux per unit accretion rate, i.e., $\mathcal{F}(r)/\dot{{\rm m}}$. Moreover, $g$ is the determinant of the three-dimensional subspace with coordinates $(t,r,\varphi)$ and is given by $\sqrt{g}=\sqrt{g_{tt}g_{rr}g_{\varphi\varphi}}$. The quantities appearing in Eq. \eqref{eq:flux} are,
\begin{eqnarray}
\Omega(r)&=& \frac{d \varphi}{dt}=\sqrt{- \frac{\partial_{r} g_{tt}}{\partial_r g_{\varphi\varphi}}}, \\
E(r)&=&u_{t}=u^{t}g_{tt}, \\
L(r)&=&-u_{\varphi}=-u^{\varphi}g_{\varphi\varphi} =-\Omega u^{t}g_{\varphi\varphi}, \\
u^{t}(r)&=&\dot{t}=\frac{1}{\sqrt{g_{tt} + \Omega^{2} g_{\varphi\varphi}}},
\end{eqnarray}
namely $\Omega=\Omega(r)$ is the orbital angular velocity, $E=E(r)$ is the energy per unit mass and $L=L(r)$ is the orbital angular momentum per unit mass of the test particle. Additionally, $\partial_{r}$ is the derivative with respect to the radial coordinate $r$, a dot represents the derivative with respect to the proper time and $u^t$ is the time component of the 4-velocity.

Another important quantity is the differential luminosity that is interpreted as the energy per unit of time reaching an observer at infinity. We denote it by $\mathcal{L}_{\infty}$ and estimate it  through the flux, $\mathcal{F}$, by means of the following relation  \cite{novikov1973, page1974}
\begin{equation}
 \label{eq:difflum}
\frac{d\mathcal{L}_{\infty}}{d\ln{r}}=4\pi r \sqrt{g}E \mathcal{F}(r).\\
\end{equation}

If the radiation emission is assumed to be well described by that of a black body, then we can express the spectral luminosity at infinity $\mathcal{L}_{\nu,\infty}$ as a function of the radiation's frequency $\nu$ as \cite{2020MNRAS.496.1115B}
\begin{equation}
 \label{eq:speclum}
\nu \mathcal{L}_{\nu,\infty}=\frac{60}{\pi^3}\int^{\infty}_{r_i}\frac{\sqrt{g }E}{M_T^2}\frac{(u^t y)^4}{\exp\left[u^t y/\mathcal{F}^{*1/4}\right]-1}dr,
\end{equation}
where $y=h\nu/kT_*$, $h$ is the Planck constant, $k$ is Boltzmann's  constant, $M_T$ is the total mass, $\mathcal{F}^*=M_T^2\mathcal{F}$ and we have taken $r_i=r_{ISCO}$. Also, $T_*$ is the characteristic temperature defined from the Stefan-Boltzmann law, which reads
\begin{equation}
\sigma T_*=\frac{\dot{{\rm m}}}{4\pi M_T^2}\,,
\end{equation}
with $\sigma$ the Stefan-Boltzmann constant.


To compare with a Schwarzschild black hole in vacuum it is also interesting to calculate the radiative efficiency of the source, i.e. the amount of rest mass energy of the disk that is converted into radiation which is given by
\be\label{eq:eff}
\mathcal{L}_\infty/\dot{{\rm m}}=1-E(r_{ISCO}),
\ee
that in the case of Schwarzschild gives the known result $\eta=(1-E(r_{ISCO}))\times 100\%\simeq 5.7\%$.

\section{Discussion of numerical results}\label{sez3}

To compute orbital parameters of test particles, flux, differential and spectral luminosities of an accretion disk, we need first to numerically solve the TOV equations. The corresponding numerical solutions for the pressure and the metric functions must fulfill the boundary conditions above reported in Eqs.~\eqref{eq:bc}-\eqref{eq:bc22}. In particular, holding  $P(r)=0$ at $r=r_s$, i.e. at the surface radius of the dark matter envelope, implies that $N$ and $\Lambda$ have to ensure $N(r_s)=-\Lambda(r_s)$ on the surface. However, the numerical value of function $N(r)$ which denote as $N_{n}(r_s)$, obtained from the numerical solution of the TOV equations, is not equal to $-\Lambda(r_s)$. This is related to the fact that the boundary condition $N(r_b)=\ln{(1-r_g/r_b)}$ is imposed while solving the TOV equations, whereas the  boundary condition $N(r_s)=\ln{(1-2M(r_s)/r_s)}$ is not. Therefore in order to satisfy the latter boundary condition one needs a redefinition of function $N_{n}$.
The most suitable redefinition of $N_n$ is the following
\begin{equation}
N_{r}(r)=N_{n}(r)-\left[N_{n}(r_s)-\ln{\left(1-\frac{2M(r_s)}{r_s}\right)}\right]\frac{r-r_b}{r_s-r_b}, \qquad
\end{equation}
which clearly fulfills the boundary conditions in Eq.~\eqref{eq:bc}.

\begin{figure*}[ht]
\begin{minipage}{0.51\linewidth}
\center{\includegraphics[width=0.97\linewidth]{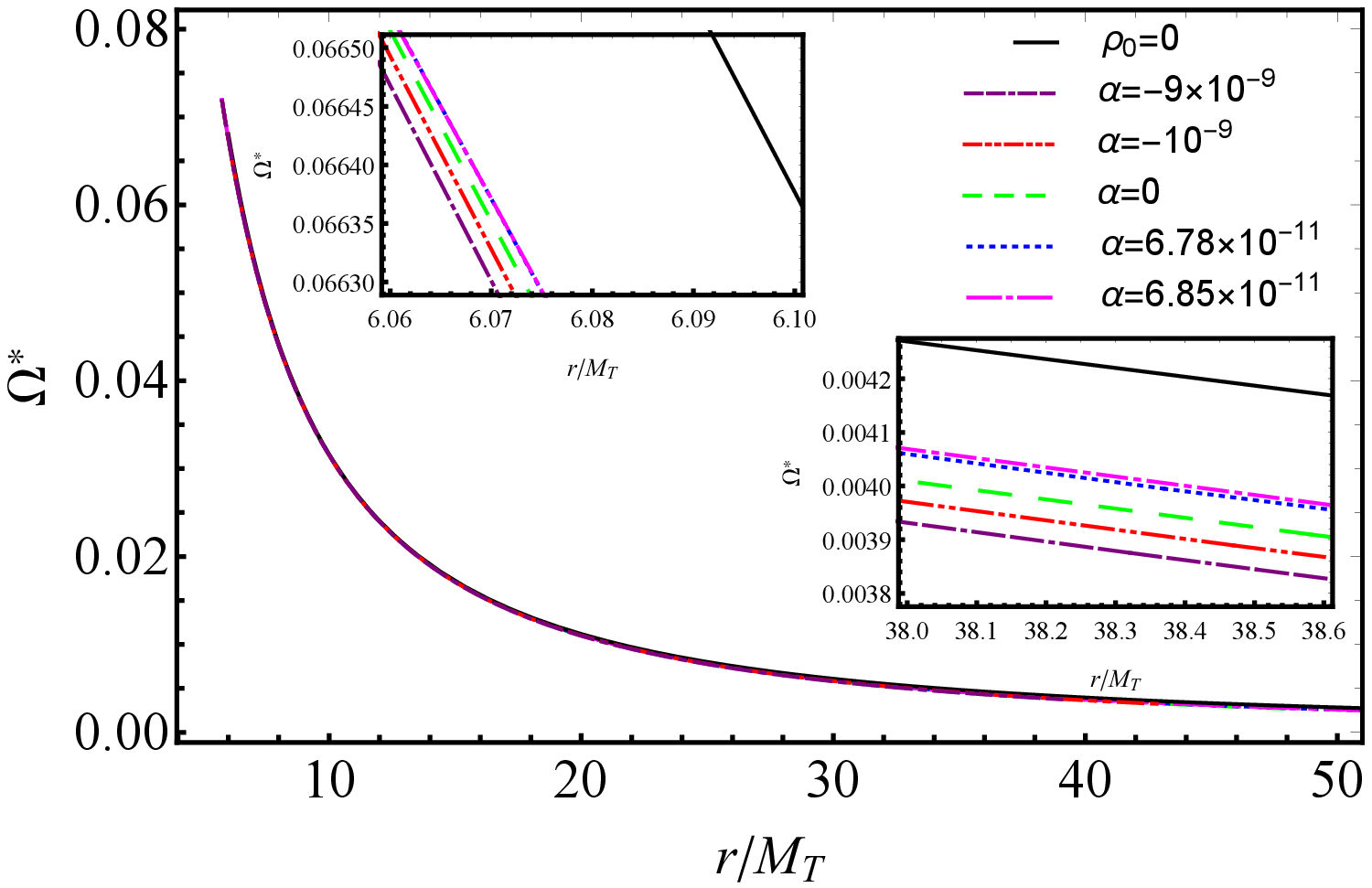}\\ }
\end{minipage}
\hfill
\begin{minipage}{0.48\linewidth}
\center{\includegraphics[width=0.97\linewidth]{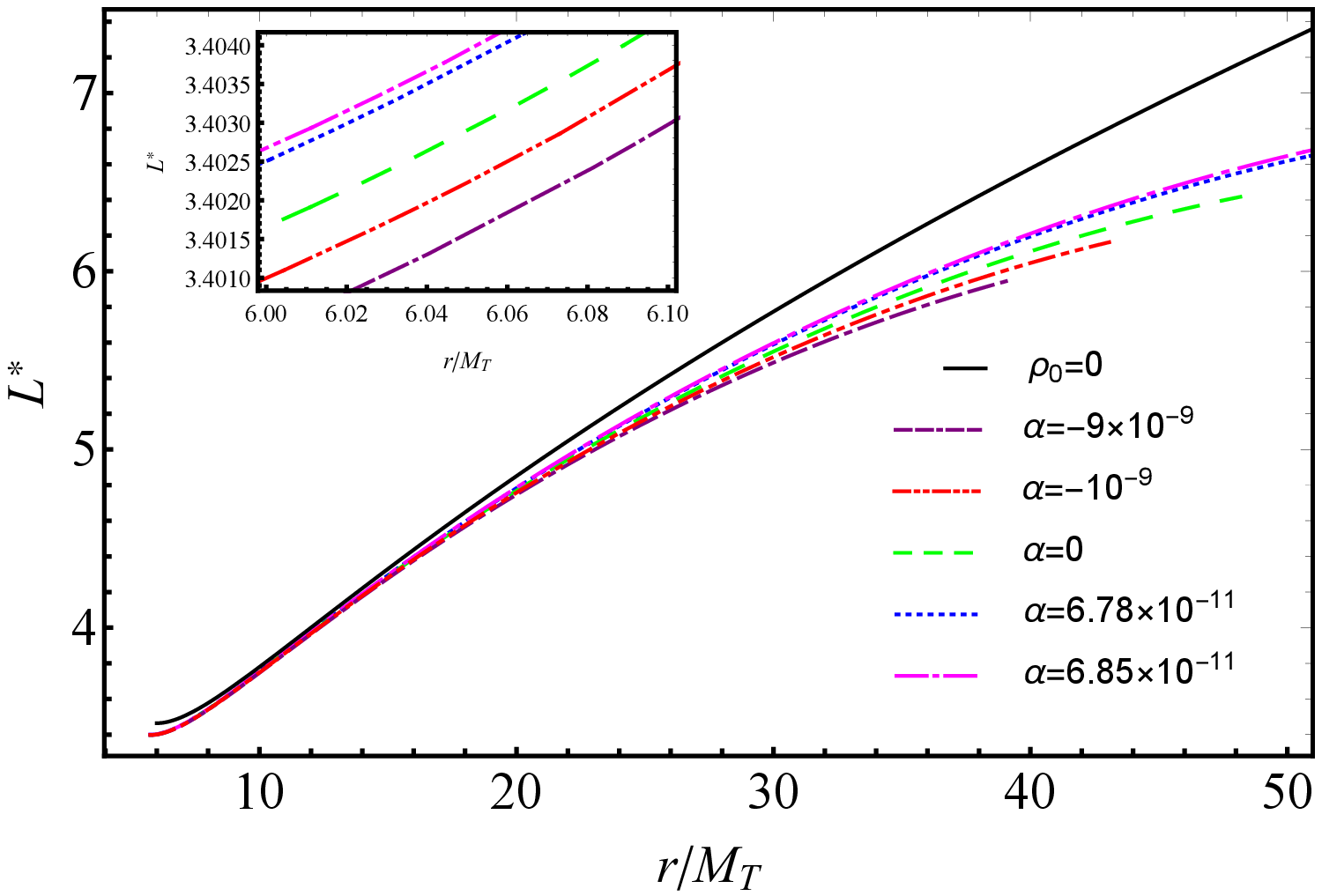}\\ }
\end{minipage}
\caption{Color online. Left panel: numerical evaluation of the orbital angular velocity $\Omega^*$ of test particles in the accretion disk around a static black hole of mass $M_{BH}=5\times 10^8 M_{\odot}\approx 4.933 AU$ in the presence of anisotropic dark matter as a function of $r/M_T$. Right panel: numerical evaluation of orbital angular momentum $L^*$ of test particles in the accretion disk as a function of $r/M_T$. In both figures the solid black curves represents the case of a static black hole without dark matter while the other curves represent anisotropic dark matter envelopes with $\rho_0=0.85\times 10^{-5} AU^{-2}$}.
\label{fig:omega and L}
\end{figure*}

\begin{figure*}[ht]
\begin{minipage}{0.49\linewidth}
\center{\includegraphics[width=0.97\linewidth]{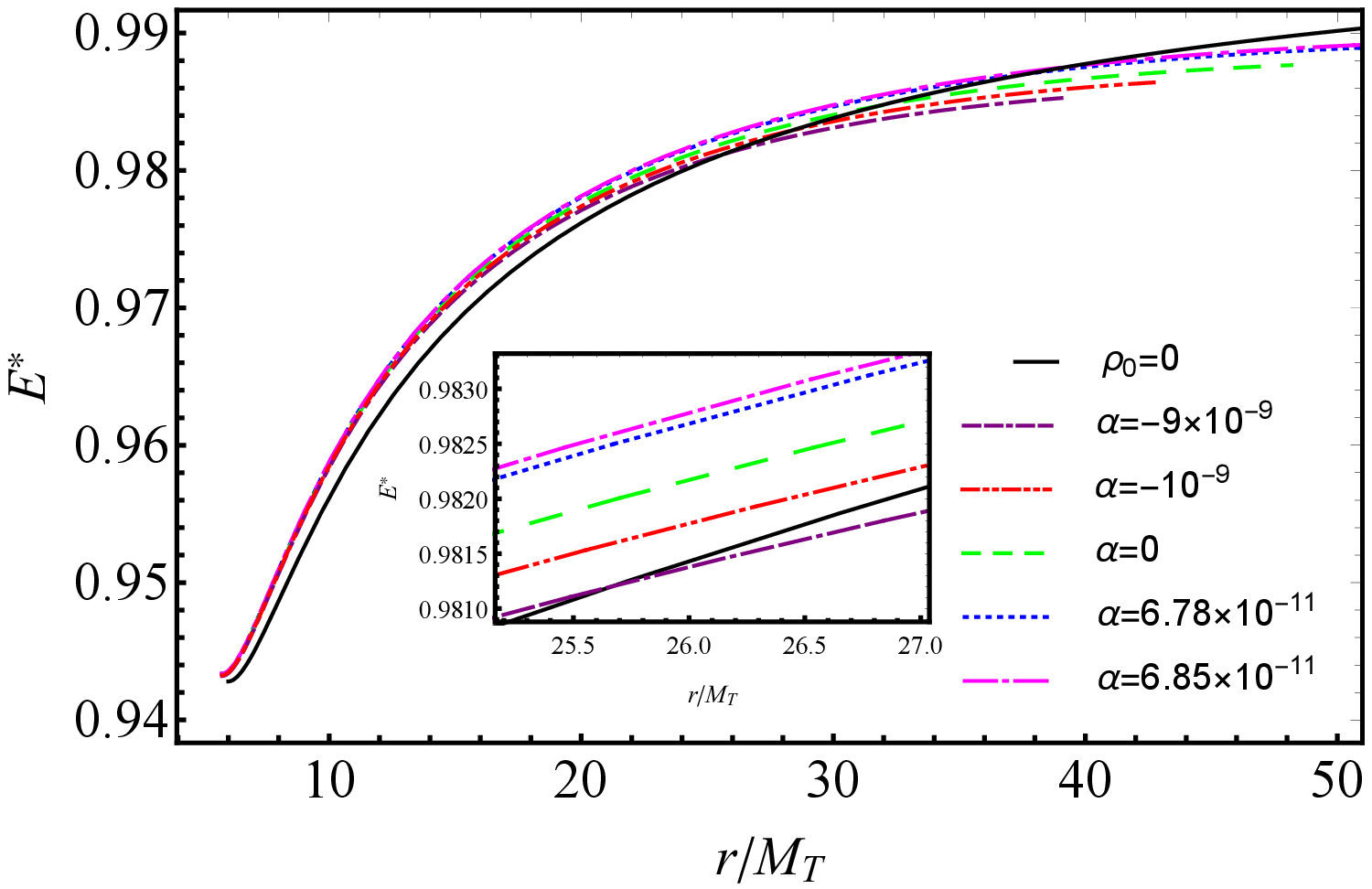}\\ }
\end{minipage}
\hfill
\begin{minipage}{0.49\linewidth}
\center{\includegraphics[width=0.97\linewidth]{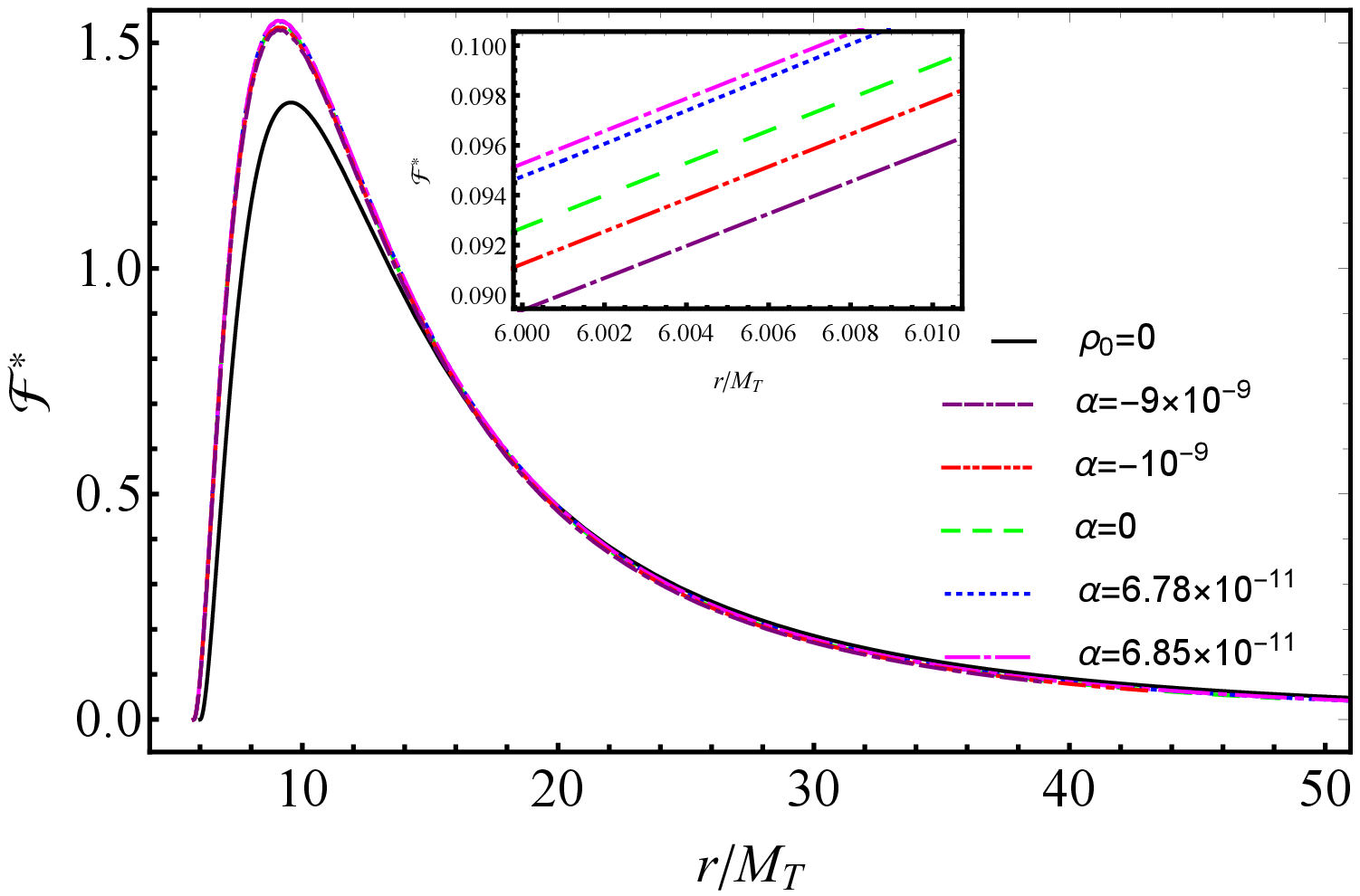}\\ }
\end{minipage}
\caption{Color online. Left panel: numerical evaluation of energies $E^*$ of test particles in the accretion disk around a static black hole of mass $M_{BH}=5\times 10^8 M_{\odot}\approx 4.933 AU$ in the presence of anisotropic dark matter as a function of $r/M_T$. Right panel: numerical evaluation of the flux $\mathcal{F}$ divided by $10^{-5}$ of the accretion disk as a function of $r/M_T$. In both figures the solid black curves represents the case of a static black hole without dark matter while the other curves represent anisotropic dark matter envelopes with $\rho_0=0.85\times 10^{-5} AU^{-2}$.}
\label{fig:E and Flux}
\end{figure*}

\begin{figure*}[ht]
\begin{minipage}{0.49\linewidth}
\center{\includegraphics[width=0.97\linewidth]{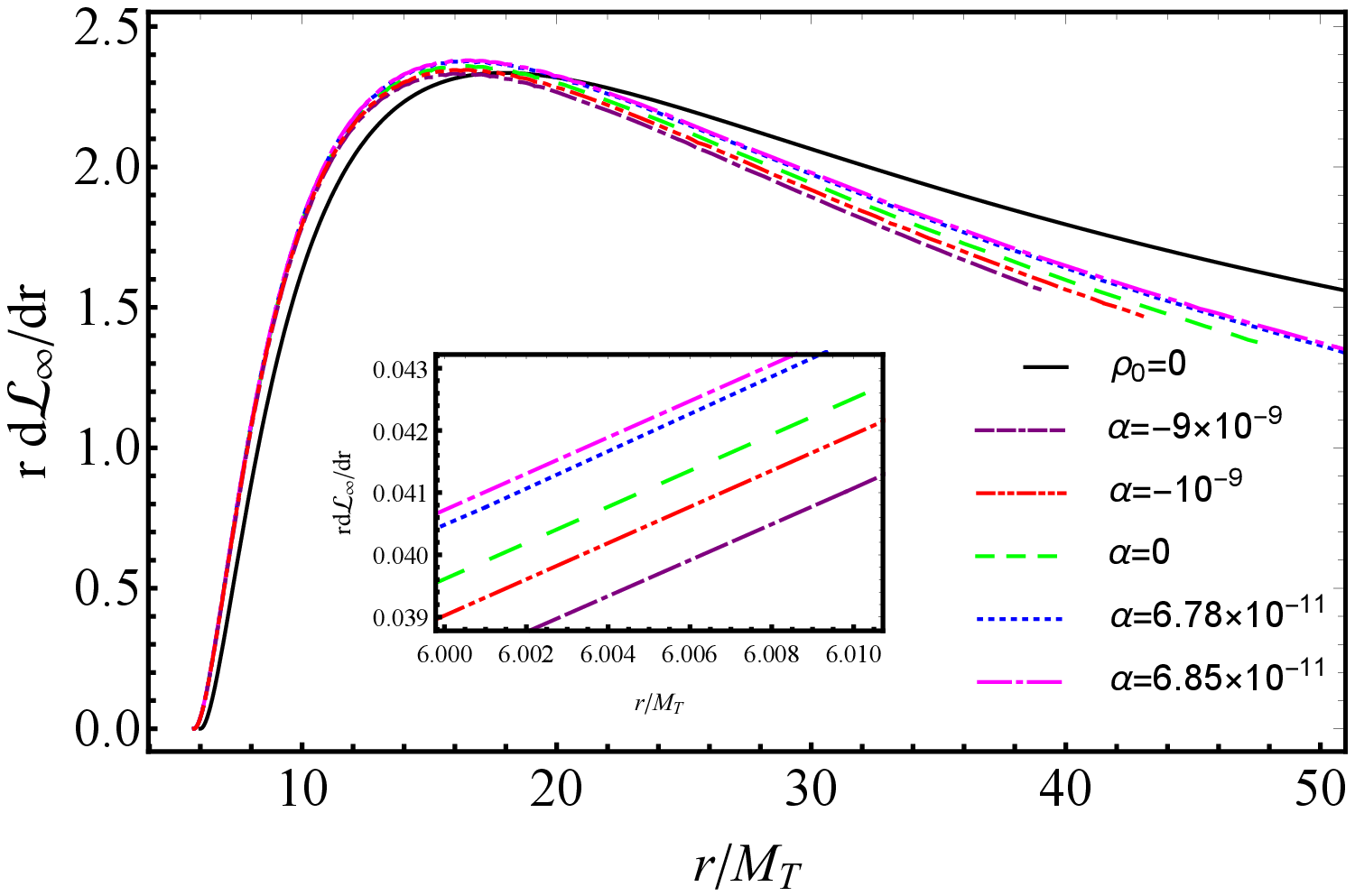}\\ }
\end{minipage}
\hfill
\begin{minipage}{0.50\linewidth}
\center{\includegraphics[width=0.97\linewidth]{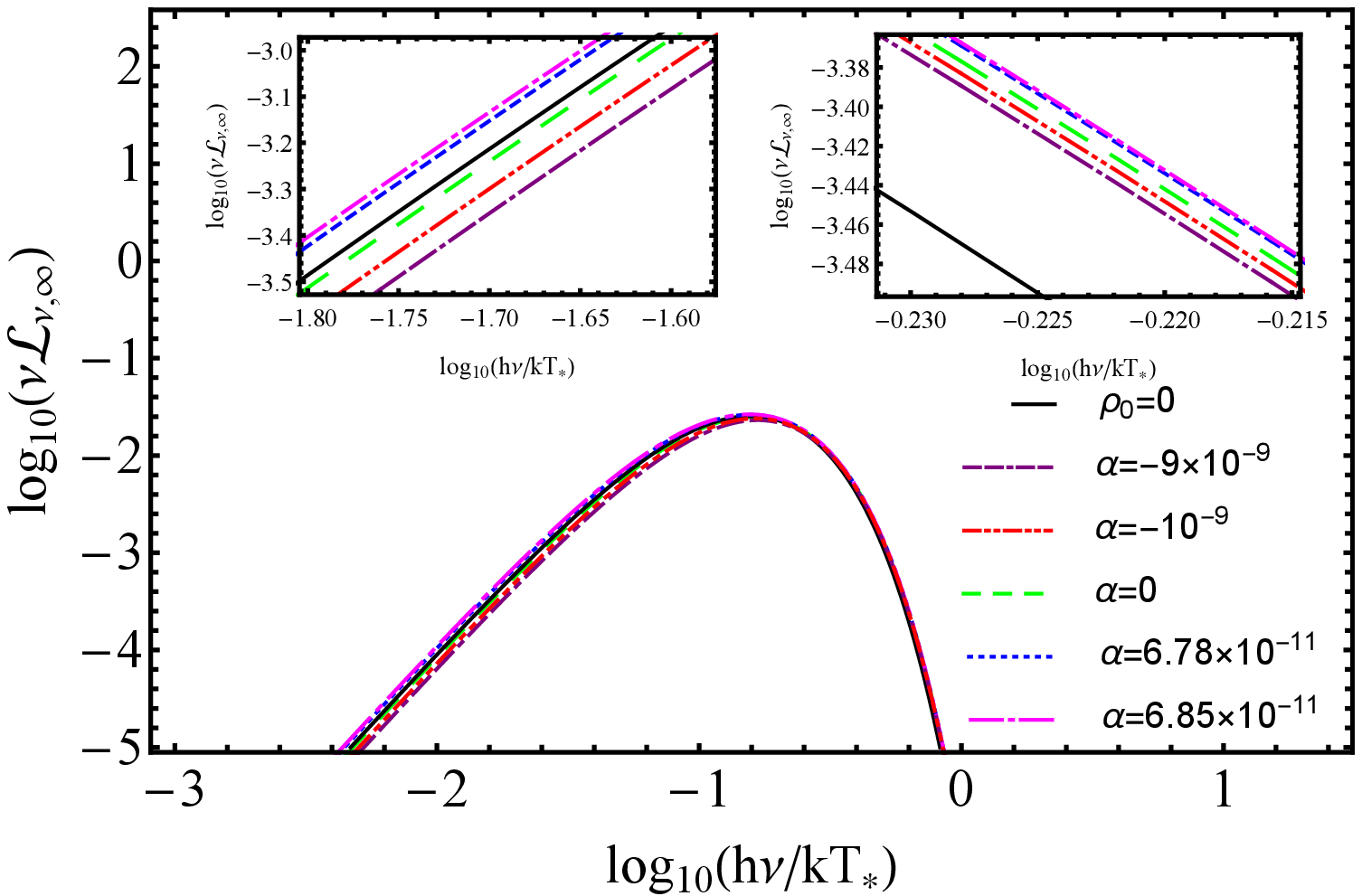}\\ }
\end{minipage}
\caption{Color online. Left panel: numerical evaluation of the differential luminosity of the accretion disk scaled in powers of $10^{-2}$ as a function of $r/M_T$. Right panel: numerical evaluation of the spectral luminosity of the accretion disk as a function of $h\nu/kT_*$ , i.e. as a function of frequency. In both figures the solid curves represents the case of a static black hole without dark matter.
The intersection points are listed in Table~\ref{tab:alpharho}.}
\label{fig:diffspecLum}
\end{figure*}
%



The numerical analysis that follows shows the effects of the presence of the dark matter envelope as depending on the value of the parameter $\alpha$ in Eq. \eqref{eq:delta}. It is worth noticing that in order to solve the TOV equations one needs to restrict $\alpha$ to negative or small positive values.
The outer boundary of the envelope $r_s$ is determined from the TOV equations by imposing $P(r_s)=0$, once $\rho_0$, $P_b$ and $\alpha$ are fixed. Finally the total amount of dark matter in the envelope is given by $M_{DM}(r_s)$.

With these ingredients we fully determine the metric functions and therefore we can study the motion of test particles within the dark matter cloud under the hypothesis that dark matter doesn't interact with the baryonic matter of the accretion disk and therefore test particles move on geodesics in the geometry produced by the dark matter envelope surrounding the black hole. The value of $r_{ISCO}$ is obtained from the evaluation of stable circular orbits within the dark matter envelope and it is set as the inner edge of the disk.

Comparison of the values of $r_s$, $M_{DM}$ and $r_{ISCO}$ for different values of $\alpha$ can be found in Tab. \ref{tab:table1}.
It is worth noticing that, for a fixed value of $\rho_0$ and $P_b$ the TOV equations can be solved for a wide range of values of $\alpha<0$, which implies $P_\theta<P$. On the other hand, for $\alpha>0$, i.e. $P_\theta>P$ there are maxima of $\alpha$ for which the TOV equations lead to unstable dark matter configurations and the TOV equations will not have solutions. Also the total dark matter mass tends to reach a maximum value of $M_{DM}(r_s)\simeq 2.12342\times 10^{-2}\,M_{BH}$ for $\alpha\simeq -10^{-8}\div10^{-9}$. Similarly, considering a central black hole of mass $M_{BH}=5\times 10^8 M_{\odot}\approx 4.933 AU$ (with $M_{\odot}$ being the mass of the Sun) the value of $\alpha$ affects only slightly the location of the ISCO, while it affects significantly the outer edge of the envelope $r_s$ i.e. $r_s$ and  $M_{DM}(r_s)$ increase with increasing $\alpha$ from negative to positive values. On the contrary, the radiative efficiency of the accretion disk decreases as $\alpha$ increases. See for Tab. \ref{tab:table1} details.


\begin{table}[ht]
\begin{center}
\caption{Physical parameters of the dark matter envelope with fixed $\rho_{0}=0.85\times 10^{-5} AU^{-2}$ and $P_{b}=2.356\times10^{-8} AU^{-2}$. First column shows various values of $\alpha$, second corresponding $r_{ISCO}$, third   $r_{s}$, fourth the total mass of dark matter $M_{DM}(r_{s})$ in units of the black hole mass with $M_{BH}=5\times 10^8 M_{\odot}\approx 4.933 AU$ and fifth the radiative efficiency of the source $\eta=(1-E(r_{ISCO}))\times100\%$.
The choice of the parameter $\alpha$ determines the innermost stable circular orbit radius $r_{ISCO}$, the thickness of the dark matter envelope $r_{s}$ and its total mass.}
\vspace{3 mm}
\label{tab:table1}
\begin{tabular}{|c|c|c|c|c|}
\hline
\hline
 $\alpha$  & $r_{ISCO}$ & $r_{s}$ & $10^{-2}M_{DM}(r_{s})$ & $\eta$\  \\
& $\left(AU\right)$ & $\left(AU\right)$ & $\left(M_{BH}\right)$& (\%)\\
\hline
\hline
  $-0.10$ & $29.412$ & $57.755$ & $1.8083265$ & $5.866$ \\
  \hline
  $-0.08$ & $29.390$ & $59.405$ & $1.8433482$ & $5.856$ \\
  \hline
  $-0.06$ & $29.363$ & $61.558$ & $1.8836587$ & $5.845$ \\
   \hline
  $-0.04$ & $29.328$ & $64.638$ & $1.9320237$ & $5.830$ \\
   \hline
  $-0.02$ & $29.273$ & $70.009$ & $1.9951764$ & $5.808$ \\
   \hline
  $-0.01$ & $29.226$ & $75.490$ & $2.0389488$ & $5.789$ \\
   \hline
  $-10^{-3}$ & $29.121$ & $94.285$ & $2.1043189$ & $5.747$ \\
   \hline
  $-10^{-5}$ & $29.043$ & $133.695$ & $2.1227230$ & $5.706$ \\
  \hline
  $-10^{-7}$ & $29.005$ & $174.751$ & $2.1234056$ & $5.686$ \\
\hline
  $-9\times10^{-9}$ & $28.992$ & $196.748$ & $2.1234221$ & $5.679$ \\
\hline  $ -10^{-9}$  & $28.982$ &  $216.981$ & $2.1234243$&$5.674$\\
\hline $0$  & $28.973$& $242.696$ & $2.1234247$ & $5.670$ \\
\hline  $6.78\times 10^{-11}$  & $28.960$ & $288.444$ & $2.1234247$ & $5.664$\\
\hline  $6.85 \times 10^{-11}$ & $28.958$ & $299.523$ & $2.1234247$ & $5.662$\\
\hline  $6.8872\times10^{-11}$ & $28.942$ & $402.399$ & $2.1234247$ & $5.655$\\
 \hline
 \hline
  \end{tabular}
  \end{center}
\end{table}

With the above numerical setup, in Fig.~\ref{fig:omega and L} we plot $\Omega^\star=M_T\Omega$ and $L^\star=L/M_T$, i.e. the dimensionless orbital angular velocity and  orbital angular momentum of test particles in the presence of a dark matter envelope, as functions of $r/M_T$. The solid curves represent the case of a static black hole without dark matter, namely $\rho_0=0$, and are easily distinguished from the other curves. For other curves the values of density and pressure are fixed as  $\rho_{0}=0.85\times 10^{-5} AU^{-2}$ and  $P_{b}=2.356\times10^{-8} AU^{-2}$, and $\alpha$ is varied.
In particular the distinction is more marked for orbital angular momentum $L^\star$, especially for large values of $r/M_T$.


Similar plots for $E^\star$ and $\mathcal F^\star$, namely the energy per unit mass of test particles and the disk's flux, are obtained in Fig.~\ref{fig:E and Flux}. In the case of energy one can see that for smaller $r/M_T$ $E^\star$ is larger than the pure vacuum case while it becomes smaller $r/M_T$ increases.
Consequently, for each $\alpha$ there is an intersection point of the $\rho_0=0$ curve with the $\rho_0=0.85\times 10^{-5} AU^{-2}$ curve which produce the same energy.  In the case of flux, it is noticeable that the presence of anisotropic dark matter with different $\alpha$ increases the maximum with respect to $\rho_0=0$ case.


The differential luminosity as a function of $r/M_T$ is reported in the left panel of Fig. \ref{fig:diffspecLum}. The numerical evaluation is scaled in powers of $10^{-2}$
and shows how the absence of dark matter  produces smaller luminosity with respect to the case with dark matter for small radii up to $r/M_T\simeq 15$, and larger luminosity for larger values of $r/M_T$. This suggests that the accretion disk in the presence of dark matter should emit more energy with respect to the accretion disk in vacuum for large frequencies, as it can be seen from the right panel of Fig. \ref{fig:diffspecLum} which
shows the spectral luminosity of the accretion disk as a function of the radiation frequency in a log-log plot. In fact, all frequency ranges for $\alpha>0$ posses larger luminosity with respect to the $\rho_0=0$ case, i.e. Schwarzschild. However for $\alpha\leq0$ the situation is different and at lower frequencies the luminosity is lower and at higher frequencies the luminosity higher than the Schwarzschild case.

It is relevant to notice that, although our approach is model-dependent since it relies on the assumptions made on compactness, see  Eq. \eqref{eq:delta}, the various possible values of $\alpha$ do not impact significantly on the physics of the disk's emission. This can be seen from the fact that the cases with anisotropic pressures are similar to each other and to the isotropic case, i.e. $\alpha=0$, when compared with the vacuum case (solid line in the plots),
suggesting that the effects on the spectrum are due mostly to the presence of dark matter rather than to the possible anisotropies.

From an experimental perspective, in principle, it would be possible to obtain information about the presence of dark matter from the spectra of accretion disks, if the other relevant quantities, such as the black hole mass and the disk's ISCO can be determined independently.
Last but not least, it is important to remember that even the observable properties of the dark matter distribution are also model dependent, since they rely on the choice made in Eq. \eqref{eq:den}.
Although the choices of different density profiles in Eq. \eqref{eq:den}, equations of state and prescriptions for the anisotropies in Eq. \eqref{eq:delta} would modify the values of quantities such as $r_{ISCO}$ and the flux of the accretion disk, we may expect that the overall qualitative features due to the presence of dark matter would remain unchanged.


The case when $\alpha=0$, or $P_\theta=P$, has been considered in \cite{2020MNRAS.496.1115B}.
In principle, one can expect that tangential pressures alone, i.e. in the absence of the radial pressures, also known as `Einstein cluster' \cite{Einstein,Gilbert,Hogan}, could also reproduce analogous results
providing \emph{de facto} a degeneracy between the different approaches. The case of vanishing radial pressures is interesting in itself as it may approximate a rotating fluid while keeping the advantage of making the equations much easier and it has been used to model the properties of dark matter halos in \cite{Bohmer}. The observational features of accretion disks in the Einstein cluster will appear in a separate article \cite{Boshkayev}.

\begin{figure}[ht]
\includegraphics[width=1\linewidth]{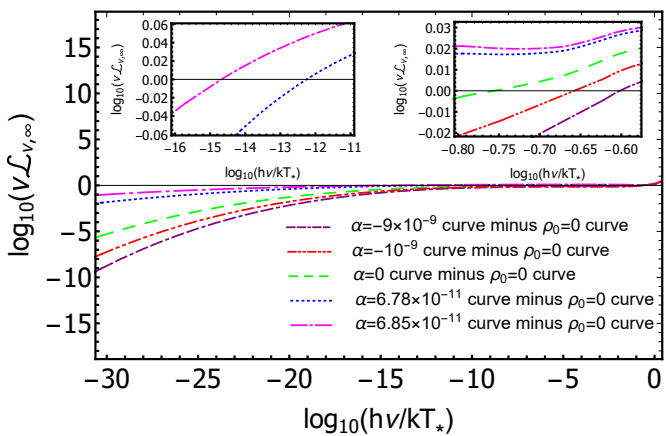}
\caption{The difference of spectral luminosity of the accretion disk between different $\alpha$ curves (with fixed $\rho_{0}=0.85\times 10^{-5} AU^{-2}$ and $P_{b}=2.356\times10^{-8} AU^{-2}$) and the vacuum case, i.e. $\rho_{0}=0$.}
\label{fig:lumdiff_alpharho}
\end{figure}

For the sake of clarity, in Fig.~\ref{fig:lumdiff_alpharho} we show the difference in luminosity between the $\rho_0=0$ (Schwarzschild) case and the cases with different values of $\alpha$ (with fixed $\rho_{0}=0.85\times 10^{-5} AU^{-2}$ and $P_{b}=2.356\times10^{-8} AU^{-2}$). The intersection points indicate that there exist frequencies for which the $\rho_0\neq0$ case with various $\alpha$ can mimic the Schwarzschild case.
However the overall spectrum for the cases in the presence of dark matter differs from that of the Schwarzschild case and the deviation, which appears to be larger at smaller frequencies, may help to constrain the values of $\alpha$ and $\rho_0$ from observations.

%

In Table \ref{fig:lumdiff_alpharho} we show numerical values of the intersection points illustrated in Fig.~\ref{fig:lumdiff_alpharho}. For fixed $\rho_{0}=0.85\times 10^{-5} AU^{-2}$ and $P_{b}=2.356\times10^{-8} AU^{-2}$ and increasing values of $\alpha$ the difference in luminosity decrease and so does the frequency. These numbers show the frequency at which the disk surrounding a Schwarzschild black hole in vacuum can be mimicked by that of a black hole surrounded by dark matter with a given value of $\alpha$.

\begin{table}[ht]
\begin{center}
\caption{The intersection points between several luminosity curves with $\rho_{0}=0.85\times 10^{-5} AU^{-2}$ and $P_{b}=2.356\times10^{-8} AU^{-2}$ and different values of $\alpha$ and the luminosity curve of a Schwarzschild black hole in vacuum $\rho_{0}=0$.
First column shows the different values of $\alpha$ considered, the second and third columns show the frequencies of the emitted radiation and the spectral luminosity as per Fig.~\ref{fig:lumdiff_alpharho}.}
\vspace{3 mm}
\label{tab:alpharho}
\begin{tabular}{|c|c|c|}
\hline
\hline
 $\alpha $ & $ \log_{10}(h\nu/kT_{*})$ &  $\log_{10}(\nu\mathcal{L}_{\nu,\infty})$  \\
\hline
\hline
  $-9\times10^{-9}$ & $-0.603$ &  $-1.764$ \\
\hline $-10^{-9}$  & $-0.651$ & $-1.670$ \\
\hline  $0$  & $-0.756$  & $-1.606$\\
\hline  $6.78\times10^{-11}$ & $-12.254$ &  $-33.397$\\
\hline  $6.85\times10^{-11}$ & $-14.683$ &  $-39.156$\\
 \hline
 \hline
  \end{tabular}
  \end{center}
\end{table}
%


\begin{figure}[ht]
\includegraphics[width=1\linewidth]{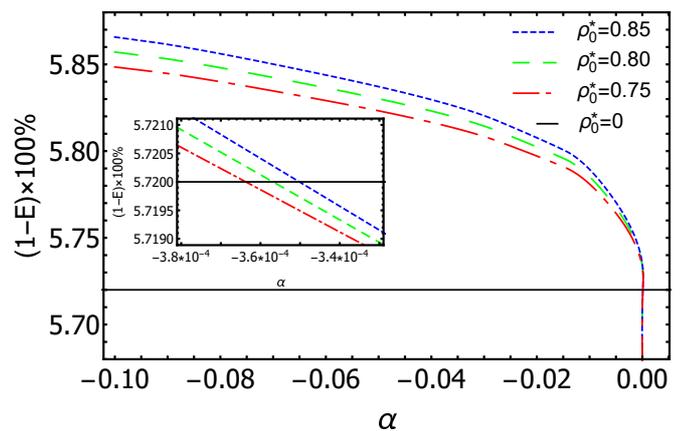}
\caption{Efficiency $\eta=(1-E(r_{ISCO}))\times100\%$ versus $\alpha$. Black solid line corresponds to the efficiency of the Schwarzschild black hole. Color curves shows efficiency in the presence of dark matter with different densities.}
\label{fig:alphaeff}
\end{figure}

Finally in Fig.~\ref{fig:alphaeff} we plot the efficiency of the source as a function of $\alpha$ for different densities $\rho_0^*=0.85$, 0.80 and 0.75, where $\rho_0^*=\rho_0/(10^{-5}{\rm AU}^{-2})$. It is evident that negative $\alpha$ will yield larger efficiency with respect to the Schwarzschild black hole in vacuum (5.72$\%$) and for  positive $\alpha$ the efficiency will be slightly smaller than 5.72$\%$. For numerical values of the efficiency see Table~\ref{tab:table1}.

\section{Final outlooks}\label{sez4}



We considered a spherically symmetric configuration composed of a central black hole surrounded by a spherical dark matter envelope with anisotropic pressures and studied the spectra produced by the accretion disk surrounding the central object in the assumption that the baryonic matter in the disk does not interact with the dark matter particles in the envelope.


Under a series of assumptions for the dark matter component of the system, namely density profile, anisotropies and inner boundary, we solved the TOV equations to determine the geometry inside the dark matter envelope and consequently the motion of test particles within the disk. We then numerically evaluated the flux and luminosity of the disk in the presence of dark matter and compared it with the isotropic case and the vacuum case.

We showed that there exist frequencies in the spectrum of accretion disk that bear the mark of the presence of dark matter and that are also affected by the anisotropies. This suggests that if other relevant quantities, such as the mass of the central black hole and the innermost stable circular orbit, can be determined independently, it could be possible, at least in principle, to distinguish different cases.
In addition, we also estimated how the radiative efficiency of the source is affected by the presence of dark matter anisotropies and found that the efficiency increases with respect to that of a Schwarzschild black hole immersed in isotropic dark matter when $P_\theta<P$.

Of course, the scenario presented here is just a simple toy model to highlight the qualitative features that the presence of anisotropies may bear on the accretion disk's spectrum. Astrophysical black holes are expected to be rotating and, therefore, the assumption of staticity, while simplifying the equations, is not particularly realistic. However, we expect similar results to hold in the presence of rotation of the central object and we aim at investigating those in future works.

At present, our knowledge of the dark matter distribution near the center of galaxies is very limited, with most studies providing estimates for the dark matter density at distances of the order of several parsecs from the galactic center in the Milky Way. Similar estimates for other galaxies are missing.
Similarly, we still don't know whether the geometry near compact objects at the center of galaxies is well described by the Kerr metric. We showed here that the presence of dark matter may affect the spectrum of the black hole's accretion disk and, therefore, provide valuable information on the nature of dark matter itself. The hope is that future observations will allow to test such ideas and further constrain the properties of viable dark matter candidates.



\begin{acknowledgments}
KB, TK, EK, OL, and HQ acknowledge the Ministry of Education and Science of the Republic of Kazakhstan, Grants: IRN AP08052311, BR05236322 and AP05133630. DM and KB acknowledge support by Nazarbayev University Faculty Development Competitive Research Grant No. 090118FD5348.

\end{acknowledgments}

\end{document}